# "Wet-to-dry" conformational transition of polymer layers grafted to nanoparticles in nanocomposite

*Chloé Chevigny[1], Jacques Jestin[1],\*, Didier Gigmes[2], Ralf Schweins[3], Emanuela Di-Cola[4], Florent Dalmas[5], Denis Bertin[2], François Boué[1]*

[1]Laboratoire Léon Brillouin, CEA Saclay 91191 Gif-sur-Yvette Cedex France

[2]Laboratoire Chimie Provence, UMR 6264, CNRS et Universités d'Aix-Marseille 1,2 et 3, Site de St Jérôme, Av. Esc. Normandie-Niemen case 542, 13393 Marseille Cedex 20, France

[3]Institut Laue Langevin DS/LSS 6 rue Jules Horowitz, 38042 Grenoble Cedex 9, France

[4]European Synchrotron Radiation Facility, 6 rue Jules Horowitz B.P. 220, 38043 Grenoble Cedex 9 France

[5]Institut de Chimie et des Matériaux Paris-Est, CNRS UMR 7182, 2-8 rue Henri Dunant 94320 Thiais France

\*Corresponding authors: Jacques.jestin@cea.fr

Abstract

The present communication reports the first direct measurement of the conformation of a polymer corona grafted around silica nano-particles dispersed inside a nanocomposite, a matrix of the same polymer. This measurement constitutes an experimental breakthrough based on a refined combination of chemical synthesis, which permits to match the contribution of the neutron silica signal inside the composite, and the use of complementary scattering methods SANS and SAXS to extract the grafted polymer layer form factor from the inter-particles silica structure factor. The modelization of the signal of the grafted polymer on nanoparticles inside the matrix and the direct comparison with the form factor of the same particles in solution show a clear-cut change of the polymer conformation from bulk to the nanocomposite: a transition from a stretched and swollen form in solution to a Gaussian conformation in the matrix followed with a compression of a factor two of the grafted corona. In the probed range, increasing the interactions between the grafted particles (by increasing the particle volume fraction) or between the grafted and the free matrix chains (decreasing the grafted-free chain length ratio) does not influence the amplitude of the grafted brush compression. This is the first direct observation of the wet-to-dry conformational transition theoretically expected to minimize the free energy of swelling of grafted chains in interaction with free matrix chains, illustrating the competition between the mixing entropy of grafted and free chains, and the elastic deformation of the grafted chains. In addition to the experimental validation of the theoretical prediction, this result constitutes a new insight for the understanding of the general problem of dispersion of nanoparticles inside a polymer matrix for the design of new nanocomposites materials.

Soft polymer properties are improved by inclusion of small hard inorganic particles inside the melt matrix. More recently, special attention has been focused on the reduction in filler size down to the nanometer range, to enhance the macroscopic properties of the nanocomposites for the design of new materials for applications in mechanical, optical, fuel cells or gas barrier engineering. The key point is the compatibility between the nano-filler and the polymer matrix, and one active field is chain grafting synthesis to dress the particle with a polymer corona of the same nature as the polymer matrix. Many recent studies present various refined synthesis processes of controlled polymerization to obtain well defined grafted nano-particles[1]. Depending on the synthesis conditions, it is possible in a controlled way to change the grafting density, the length of the grafted chains or the particles colloidal stability. The resulting grafted objects can be used as fillers by mixing with free chains of the same polymer to create new hierarchical structures inside the matrix, from well dispersed state to large anisotropic or fractal aggregates without direct connectivity, or with formation of a continuous network[2]. Theoretical interactions have been described for planar[3,4] or curved surfaces[5,6,7]. The free energy of swelling can be expressed by two terms: a mixing entropy term between the grafted and the free chains and an elastic term of deformation (stretching or compression) of the grafted chains. The low mass free entities (solvent molecules, monomer or polymer chains corresponding to a grafted-free chain length ratio superior to one) can penetrate into the grafted corona according to a favorable entropic potential and swell it: the grafted chains are stretched and the corona is "wet". Conversely, when grafted and free chain masses are comparable (grafted-free chain length ratio equal or inferior to one), mixing entropy is much lower, swelling elastic energy dominates, which expels free entities from the corona and changes its conformation. As a consequence, the grafted chains are compressed and the corona becomes "dry". These concepts have been studied experimentally in the case of hydrogels[8-9] or large silica particles[10] through indirect measurements of the inter-particles structure factor with X-rays or light scattering. But a direct measurement of the grafted polymer conformation and its change due to specific interactions with free chains is still missing, particularly in a nanocomposite system in which it is technically difficult to distinguish the grafted chains from the free ones.

The present communication reports the first direct measurement of the conformation of the polymer brush grafted around silica nano-particles inside a matrix of the same polymer. This success is based on two main contributions. First, the development of refined chemistry to build controlled grafted silica particles with deuterated polystyrene (PS) chains and to synthesize a statistical hydrogenated-deuterated polystyrene matrix which matches the particles scattering in Small Angles Neutron Scattering (SANS) experiments: the silica becomes invisible to neutrons in the matrix and only the deuterated corona contributes to the signal. Secondly, a combination of Small Angles and Ultra Small Angles X-rays Scattering (SAXS/USAXS) to extract the inter-particles structure factor, corresponding to the organization of the grafted particles inside the polymer matrix with microscopy (TEM) to check the particle dispersion in real space.

We use a "grafting from" technique associated with a radical-controlled polymerization (Nitroxyde-Mediated-Polymerization, NMP) to obtain well defined silica particles (Ludox type) grafted with polystyrene chains. The grafting density is fixed at 0.2 molecules/nm² and the polymerization gives chains of $M_n$ = 25 000 g/mol with low poly-dispersity index around 1.2. The complete description of the grafting process can be founded in reference 11. The synthesis process permit to graft either normal or deuterated polymers at the surface of the particles but only deuterated grafted chains were used in the present paper. The grafted corona can be modelized in solution using SANS contrast variation method with a core-shell model from which we can extract a layer thickness[11]. We have then well defined grafted nano-particles, which can be suspended in an organic solvent (Dimethylacetamide, DMAc), and mixed with PS chains at 10% v/v dissolved in the same solvent. Final nanocomposites are formed according to a controlled solvent slow evaporation process described previously[12] which permits to obtain homogenously filled films samples. The polymer chains for the matrix were synthesized by "classical" Radicalar Polymerization (6hours at 80°C) by mixing normal and deuterated styrene to form a statistical H/D PS chain at the final state of the polymerization process. The styrene H/D ratio was varied (from 35 to 100% v/v of H-monomer) and the neutron scattering length density (SLD) of the resulting chain measured with SANS (PAXY at LLB) to find the SLD value the closest possible of the one of the silica ($3.40 \ 10^{-6} \ Å^{-2}$) (Figure 1a). To check this, a "blank" filled with bare silica is used: the scattering coming from the silica is reduced by more than two orders of magnitude (Figure 1b). Varying the polymerization time allows to synthesize three different masses for the matrix, respectively of $M_n$ = 35 000, 58000 and 98000 g/mol.

Different nanocomposites were then formed either at fixed grafted/free matrix chain ratio ($M_{n \ grafted}$ = 25000 and $M_{n \ matrix}$ = 98000 g/mol) with growing quantities of grafted silica particles (from 5 to 20% v/v) or at fixed particles content (5% v/v) by varying the chain length ratio ($M_{n \ grafted}$ = 25 000 and $M_{n \ matrix}$ = 35 000, 58000 and 98000 g/mol). The samples were analyzed with a combination of SAXS and USAXS measurements on ID02 at ESRF using the classical pinhole setup and the Bönse-hart setup. Results are presented on figure 2, (a) for concentration variation and (b) for chain length ratio variation. As X-rays do not differentiate between labeled polymers, these spectrums reveal the silica particles dispersion inside the polymer matrix related to local organization of the particles between themselves, characterized by the inter-particles structure factor S(Q). The strong upturn in the very low Q range comes from the formation of void cavities inside the materials related to the processing formation used for the films[13]. According to the general scattering equation of a particle system in a continuous medium $I_{SAXS}(Q) \sim A.S_p(Q)*P_p(Q)$ (A is proportional to the product of the contrast term $\Delta\rho^2$, the volume of the scattered particle and the number of scattered objects, the index p represent the silica particles), the inter-particles structure factor can be calculated by dividing the total scattered intensity, I(Q), by the single form factor of the silica particle P(Q). The form factor of the silica particles is well known in our case as it can be determined easily through the analysis of the scattering of a diluted solution of particles (see the full grey line in figure 2(a)). The silica particles form factor follows the classical behavior of poly-disperse (log-normal) spheres of a mean radius of <R>=13 nm with a dispersion of 0.14. The inter-particles structure factors deduced from the division of the intensity by the silica form factor are presented in figure 2(c) for the evolution as a function of the particle concentration and in figure 2(d) for the evolution as a function of the chain length ratio. The appearance of a strong correlation peak indicates a well dispersed repulsive organization of the particle inside the polymer matrix with a first neighboring typical distance equal to $2\pi/Q^*$ ($Q^*$ is the position of the first maximum peak). The evolution of the peak position $Q^*$ with the silica concentration (figure 2(c)) follows the classical behavior of a 3D cubic volume system while the peak position does not move as a function of the chain length ratio (figure 2(d)) indicating that the dispersion of the particle does not change in the probed chain length ratio range. The good dispersion of the particles is confirmed in real space by the microscopy pictures presented in figure 3 for two particle concentrations (respectively 5% v/v figure 3(a) and 14% v/v figure 3(b)).

The same samples were then measured by SANS (D11 at ILL) and the scattering curves are presented in figure 4, (a) as a function of the grafted silica concentration and (b) as a function of the chain length ratio. The scattered intensity is thus only due to the grafted deuterated chains around the silica particles as the silica signal contribution is suppressed by the matrix one and can be expressed in this case as $I_{SANS}(Q) \sim A'S_{g\_p}(Q)*P_{g\_p}(Q)$ where the index $_{g\_p}$ represent the grafted particles. For a correct modelling of the grafted brush ($P_{g\_p}(Q)$), it is needed to suppress the contribution of the inter-particles structure factor ($S_{g\_p}(Q)$) related to the arrangement of the grafted coronas between them inside the matrix. As this organization is built across the network center of masses of the scattered grafted objects, the inter-particles structure factor is independent of the nature of the probe, i.e. is the same for neutron and for X-rays, $S_p(Q)=S_{g\_p}(Q)$. Thus, the grafted corona form factor can be extracted by dividing the SANS signal (figure 4(a) and 4(b)) by the particles structure factor deduced from SAXS (figures 2(c) and 2(d)). The results are presented in figure 4(c) for the evolution as a function of the particles content and 4(d) for the evolution as a function of the chain length ratio. The figure 4(c) shows a nice superimposition of the different curves, illustrating that the form factor of the grafted corona does not depend on the silica concentration. In other words, increasing the grafted inter-particles interactions by decreasing the mean distance between themselves does not change the conformation of the grafted brushes. In addition, the figure 4(d) shows also a nice superimposition of the different curves. Changing the interaction between the grafted and the free matrix chains is expected to have an influence on the brush conformation: reducing the free chain masses from 98 000 to 35 000 was supposed to reduce the mixing entropy term and induce an increase of the stretching rate of the grafted brushes. The grafted corona form factor can be perfectly modeled (figure 5(a)) with a Gaussian chain model[14] from which we can extract the physical parameters describing the system: the number of grafted silica particles (5% v/v), the number of polymer chains per particle (~400), the mass ($M_n$=25 000 g/mol) and the radius of gyration of grafted chains equal to 3.1 nm, corresponding to a corona thickness of 6.2 nm. The conformation of the grafted corona inside the composite is compared with the conformation of the same grafted corona in solution, before the mixing with the matrix chains. For such a solution, matching the silica signal is easy using the appropriate fractions of deuterated and hydrogenated solvent. The result, presented in figure 5(b) already published in reference 11, is perfectly fitted by a core-shell model with a homogenous corona of thickness of 12 nm.

We thus observe a clear-cut change of the grafted corona conformation from the solution to the nanocomposites: the grafted chains are stretched in solution while in the nanocomposites they return to a less expanded Gaussian coil conformation. The observation is in very good agreement with the theoretically expected wet-to-dry transition of the grafted corona conformation induced by the mixing entropy potential between the free and the grafted chains. In addition, we demonstrate that interactions with longer free chains induce a compression of the grafted chains of a factor two in thickness, to minimize the corresponding free energy. No effect of the inter-particles distance or of the variation of the grafted/free chain length ratio has been observed on the amplitude of the elastic compression of the corona. But such effect is expected to be observed for larger parameters: for higher silica volume fraction or large differences in chain length ratios. This measurement and modeling of the polymer corona grafted on particles inside a polymer matrix is thus an experimental breakthrough thanks to a combination of pertinent chemical synthesis and of complementary scattering techniques SANS and SAXS, and is important on two levels: it validates the theoretical predictions and it enables better understanding of the dispersion mechanisms of nanoparticles in nanocomposites and specifically the contribution of the brush conformation modifications in these mechanisms.

ACKNOWLEDGMENT. The authors thank CEA for the PhD Grant of C. Chevigny, A. Lapp (LLB), P. Lindner (ILL) and T. Naranayan (ESRF) for beam-time allocation.

REFERENCES

(1) Radhakrishnan, B.; Ranjan, R.; Brittain, W. J. *Soft Matter* **2006**, 2, 386-396.

(2) Akcora, P.; Liu, H.; Kumar, S. K.; Moll, J.; Li, Y.; Benicewicz, B. C.; Schadler, L. S.; Acehan, D.; Panagiotopoulos, A. Z.; Pryamitsyn, V.; Ganesan, V.; Ilavsky, J.; Thiyagarajan, P.; Colby, R. H.; Douglas, J. *Nature Materials* **2009**, 8, 354-359.

(3) De Gennes, P. G. *Macromolecules* **1980**, 13, 1069-1075.

(4) Maas, J.H.; Fleer, G. J.; Leemakers, F. A. M.; Cohen Stuart, M. A. *Langmuir* **2002**, 18, 8871-8880.

(5) Gast, A.; Leibler, L. *Macromolecules* **1986**, 19, 686-691.

(6) Leibler, L.; Pincus, P. A. *Macromolecules* **1984**, 17, 2922-2924.

(7) Gast, A.; Leibler, L. *J. Phys. Chem.* **1985**, 89, 3947.

(8) Lindenblatt, G.; Schärtl, W.; Pakula, T.; Schmidt, M. *Macromolecules* **2001**, 34, 1730-1736.

(9) Lindenblatt, G.; Schärtl, W.; Pakula, T.; Schmidt, M. *Macromolecules* **2000**, 33, 9340-9347.

(10) Green, D. L.; Mewis, J. *Langmuir* **2006**, 22, 9546.

(11) Chevigny, C.; Gigmes, D.; Bertin, D.; Jestin, J.; Boué, F. *Soft Matter* **2009**, 5 (19), 3741-3753.

(12) Jouault, N.; Vallat, P.; Dalmas, F.; Said, S.; Jestin, J.; Boué, F. *Macromolecules* **2009**, 42 (6), 2031-2040.

(13) Rottler, J. ; Robbins, M. O. *Physical Review E* **2003**, 68 :011801.

(14) Pedersen, J. S.; Gerstenberg, M. C. *Macromolecules* **1996**, 29, 1363.

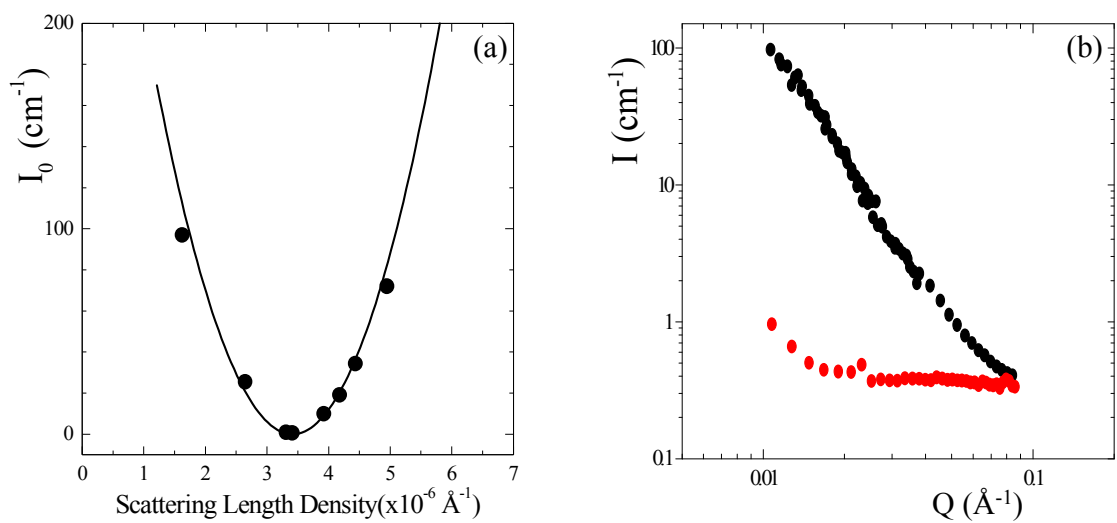

**Figure 1:** (a) variation of the H/D styrene ratio for synthesis of a statistical H/D PS matrix ($M_n$ = 98 000 g/mol) of tunable scattering length densities (SLD): the minimum of the curve correspond to the SLD of the silica, (b) Two nanocomposites filled with non-grafted silica for different SLD matrix values, far from the silica SLD point (black circles) and close to the silica SLD point (red circles).

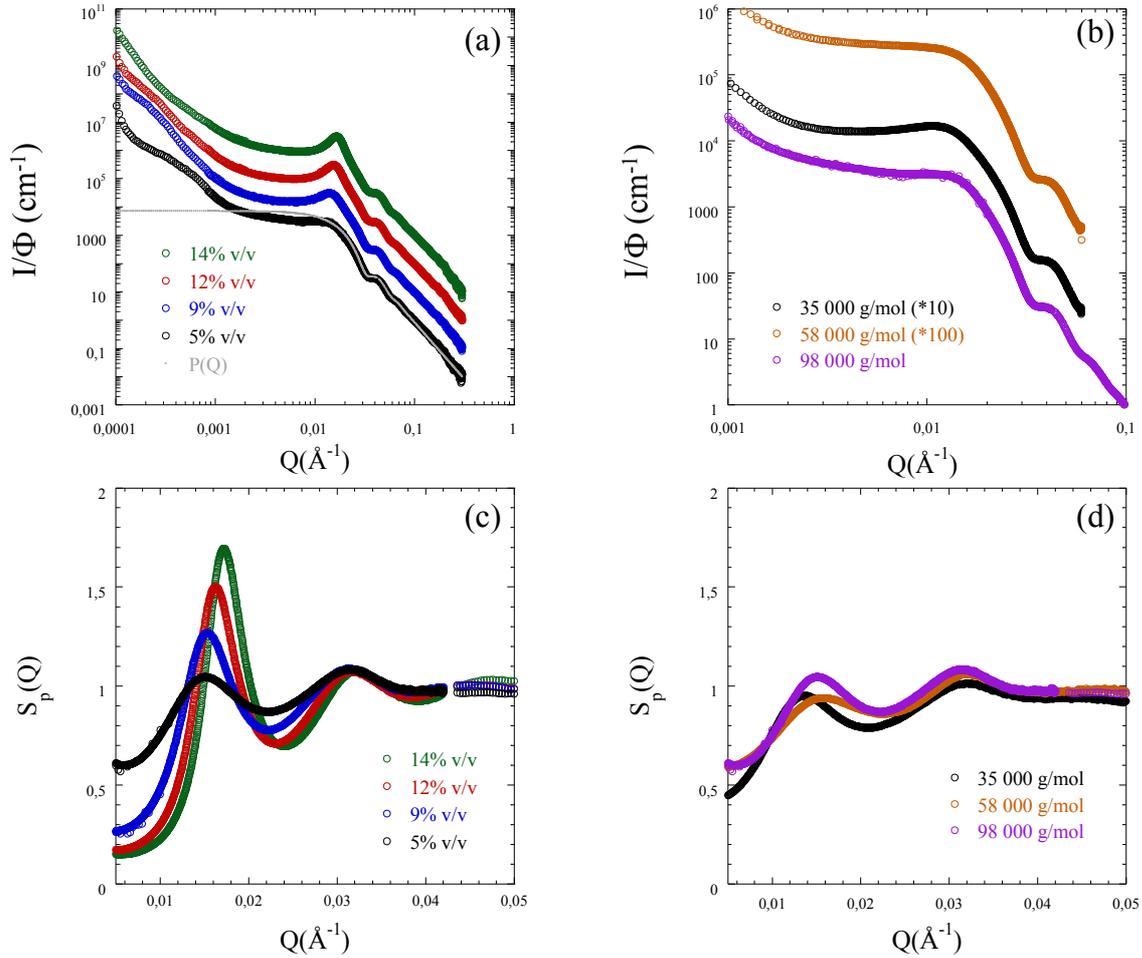

**Figure 2:** SAXS and USAXS normalized intensities (ID02 at ESRF) on PS nanocomposites filled with deuterated PS grafted silica particles as a function of (a) the silica volume fraction ranging from 5% to 14% v/v at fixed chain length ratio ($M_{n\ grafted}$ = 25 000 g/mol and $M_{n\ matrix}$ = 98 000 g/ mol) and (b) the chain length ratio at fixed silica volume fraction (5% v/v) for $M_{n\ grafted}$ = 25 000 g/mol and $M_{n\ matrix}$ = 35 000, 58 000 and 98 000 g/mol. The full gray line in (a) represent the poly-disperse sphere form factor of the native silica particle $P_p(Q)$; (c) and (d) the corresponding inter-particle structure factor $S_p(Q)$ inside the polymer matrix obtained by divided the total intensity (a) and (b) by the form factor of a single particle (full grey line).

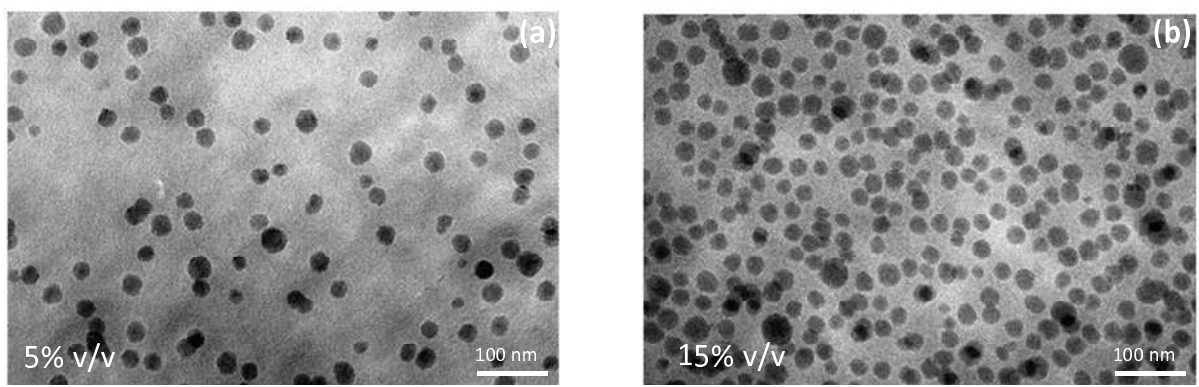

**Figure 3:** Transmission Electronic Microscopy (TEM) on the nanocomposite ($M_{n\ matrix}$ = 98 000 g/mol) filled with 5% v/v (a) and 14% v/v (b) of grafted silica particles ($M_{n\ grafted}$ = 25 000 g/mol).

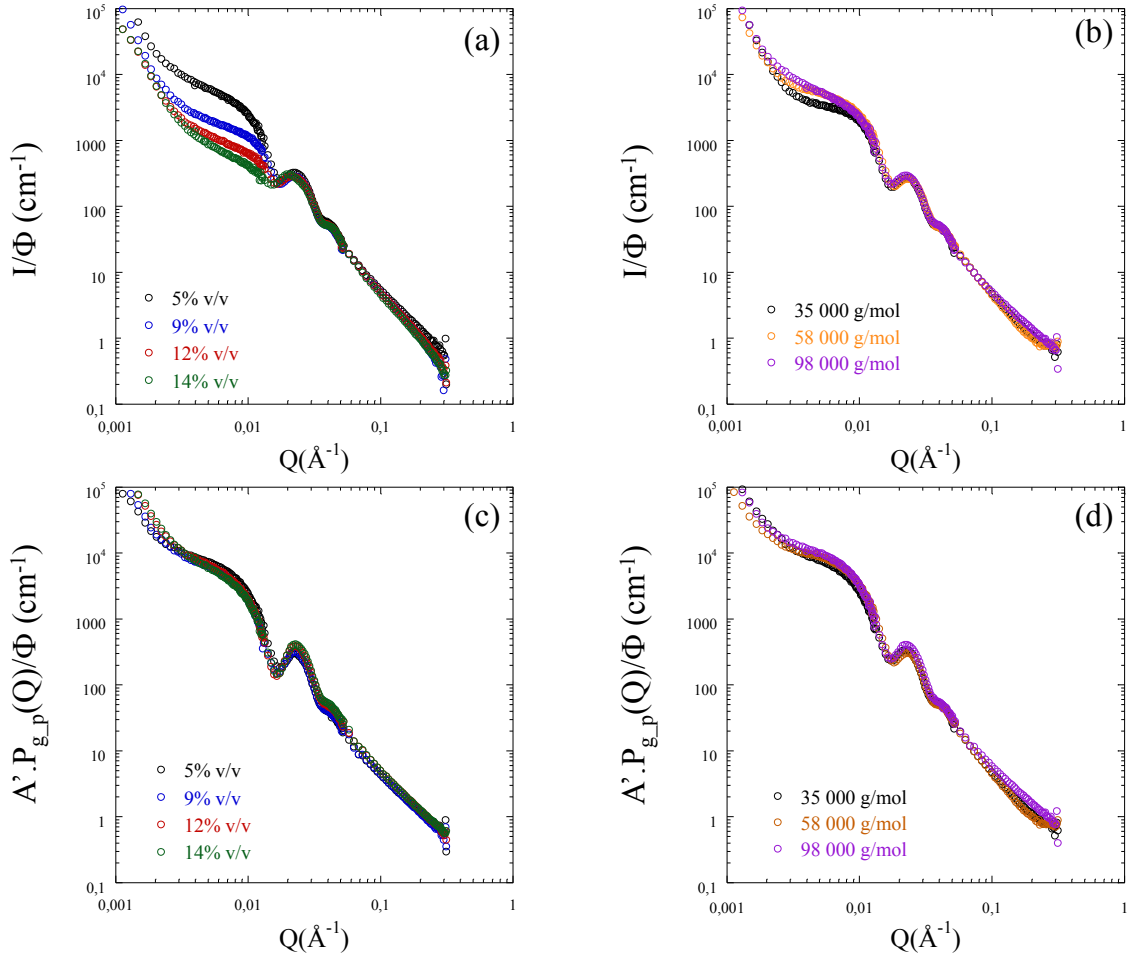

**Figure 4:** (a) SANS curves of the deuterated grafted corona ($M_{n\ grafted}$ = 25 000 g/mol) inside the silica matched matrix ($M_{n\ matrix}$ = 98 000 g/mol) at four particle concentration (5, 9, 12 and 14 % v/v), (b) SANS curves of the deuterated grafted corona ($M_{n\ grafted}$ = 25 000 g/mol) at silica volume fraction of 5% v/v inside the three silica matched matrix ($M_{n\ matrix}$ = 35 000, 58 000 and 98 000 g/mol); (c) Evolution of the form factor of the grafted polymer corona in the nano-composite $P_{p\_g}(Q)$ as a function of the silica content obtained after division of the neutron scattering intensity (figure 4(a)) by the inter-particles structure factor $S_p(Q)$ (figure 2 (c)), (d) Evolution of the form factor of the grafted polymer corona in the nano-composite $P_{p\_g}(Q)$ as a function of the chain length ratio obtained after division of the neutron scattering intensity (figure 4(b)) by the inter-particles structure factor $S_p(Q)$ (figure 2 (d)).

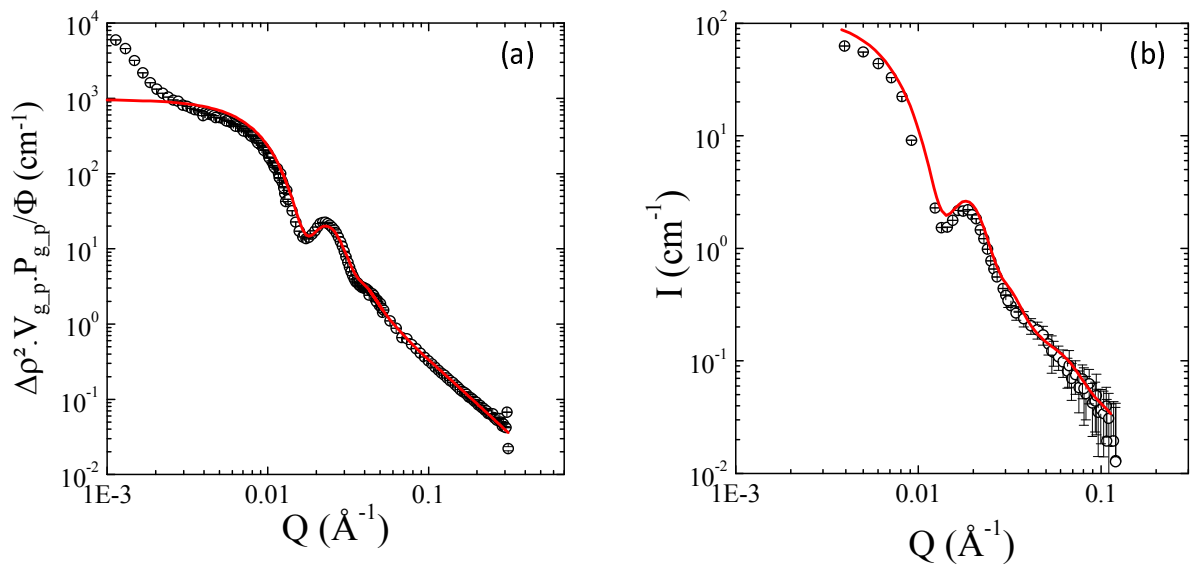

**Figure 5:** (a) modelization of the form factor of the deuterated PS grafted silica nanoparticles ($M_{n\ grafted}$ = 25 000 g/mol) dispersed in a silica matched matrix ($M_{n\ matrix}$ = 98 000 g/mol) with a Gaussian chain model for comparison with the modelization of the same grafted particles dispersed in a solvent (b) and modelized with a core-shell model (from reference 11).

TOC

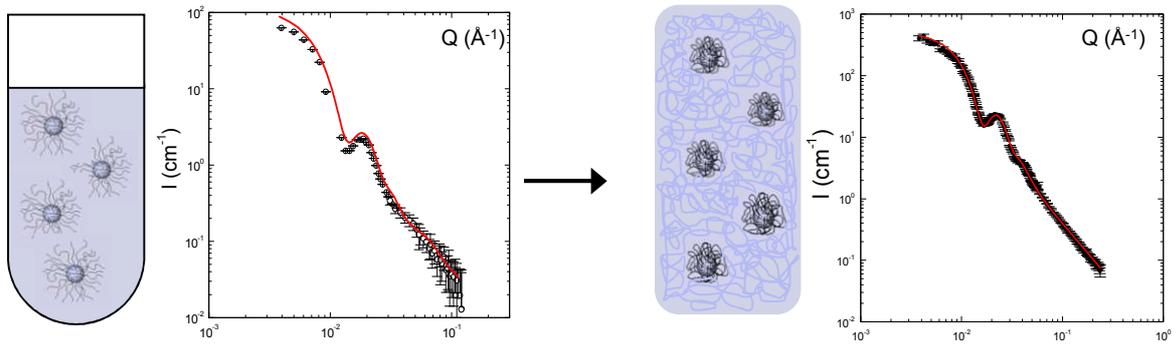

« Wet-to-dry » conformational transition of the grafted polymer layer from solution to nanocomposite